\documentclass[sigconf]{acmart}

\usepackage{multirow}
\usepackage{booktabs}
\usepackage{siunitx}

\AtBeginDocument{%
  }

\setcopyright{cc}
\setcctype{by}
\copyrightyear{2026}
\acmYear{2026}
\acmConference[SIGIR '26]{Proceedings of the 49th International ACM SIGIR Conference on Research and Development in Information Retrieval}{July 20--24, 2026}{Melbourne, VIC, Australia.}
\acmBooktitle{Proceedings of the 49th International ACM SIGIR Conference on Research and Development in Information Retrieval (SIGIR '26), July 20--24, 2026, Melbourne, VIC, Australia}
\acmISBN{979-8-4007-2599-9/2026/07}
\acmDOI{10.1145/3805712.3809877}

\begin{document}

\title{Mitigating Collaborative Semantic ID Staleness in Generative Retrieval}



\author{Vladimir Baikalov}
\email{deadinside@itmo.ru}
\orcid{0009-0009-4864-2305}
\affiliation{%
  \institution{AI VK}
  \city{Moscow}
  \country{Russia}
}
\affiliation{%
  \institution{ITMO University}
  \city{St. Petersburg}
  \country{Russia}
}

\author{Iskander Bagautdinov}
\email{iskbaga@itmo.ru}
\orcid{0009-0006-2446-7125}
\affiliation{%
  \institution{ITMO University}
  \city{St. Petersburg}
  \country{Russia}
}

\author{Sergey Muravyov}
\email{smuravyov@itmo.ru}
\orcid{0000-0002-4251-1744}
\affiliation{%
  \institution{ITMO University}
  \city{St. Petersburg}
  \country{Russia}
}

\renewcommand{\shortauthors}{Baikalov et al.}

\begin{abstract}
Generative retrieval with Semantic IDs (SIDs) assigns each item a discrete identifier and treats retrieval as a sequence generation problem rather than a nearest-neighbor search.
While content-only SIDs are stable, they do not take into account user-item interaction patterns, so recent systems construct interaction-informed SIDs. However, as interaction patterns drift over time, these identifiers become stale, i.e., their collaborative semantics no longer match recent logs.

Prior work typically assumes a fixed SID vocabulary during fine-tuning, or treats SID refresh as a full rebuild that requires retraining.
However, SID staleness under temporal drift is rarely analyzed explicitly.
To bridge this gap, we study SID staleness under strict chronological evaluation and propose a lightweight, model-agnostic SID alignment update.

Given refreshed SIDs derived from recent logs, we align them to the existing SID vocabulary so the retriever checkpoint remains compatible, enabling standard warm-start fine-tuning without a full rebuild-and-retrain pipeline. Across three public benchmarks, our update consistently improves Recall@K and nDCG@K at high cutoffs over naive fine-tuning with stale SIDs and reduces retriever-training compute by approximately 8--9 times compared to full retraining.
\end{abstract}




\maketitle

\section{Introduction}

Large-scale recommender and search systems rely on retrieval to select a small candidate set from massive catalogs~\cite{youtubednn, borisyuk2024linrmodelbasedneural, agarwal2025pinrecoutcomeconditionedmultitokengenerative}.
Generative retrieval with Semantic IDs (SIDs) offers an alternative to embedding-based retrieval.
Instead of retrieving items via vector similarity, each item is assigned a short token sequence, which the model generates from the user context~\cite{rajput2023recommendersystemsgenerativeretrieval}.

A key design choice is how to construct SIDs.
Content-only SIDs (derived from metadata or multimodal features) are stable and effective in cold-start scenarios~\cite{singh2024bettergeneralizationsemanticids}, but can miss behavioral structure captured in interaction logs~\cite{wang2025learnableitemtokenizationgenerative}.
Recent systems therefore build interaction-informed SIDs, often improving retrieval quality~\cite{he2025plumadaptingpretrainedlanguage, deng2025onerecunifyingretrieverank}.

In practical settings, collaborative patterns are non-stationary: user interests, item popularity, and logging policies evolve over time, motivating periodic SID refreshes to capture the latest collaborative structure. 
Yet refreshing SIDs introduces a challenge: new token assignments may become incompatible with the retriever’s previously learned output space. Despite its practical relevance, this issue has received limited attention in prior works.
As a result, practitioners typically face a trade-off: (i) keep SIDs fixed and finetune on new logs, accumulating SID--behavior mismatch, or (ii) refresh SIDs and retrain from scratch at high compute cost~\cite{he2025plumadaptingpretrainedlanguage, deng2025onerecunifyingretrieverank}.

However, SID staleness under temporal drift has not been systematically quantified.
To bridge this gap, we quantify interaction-informed SID staleness under strict chronological evaluation and propose a lightweight, model-agnostic alignment update that keeps refreshed identifiers checkpoint-compatible.
Our contributions are:

\begin{itemize}
    \item We quantify the degradation caused by stale interaction-informed SIDs under temporal drift.
    \item We propose a lightweight, model-agnostic SID alignment update that enables warm-start finetuning with refreshed identifiers.
    \item On three public temporal benchmarks, we show that alignment improves over stale-SID finetuning and is competitive with full retraining while reducing retriever training compute by up to 8--9 times.
\end{itemize}
\section{Related Work}

\textbf{Generative Retrieval.}
Generative retrieval frames candidate generation as sequence modeling, where a model directly decodes item identifiers~\cite{decao2021autoregressiveentityretrieval, tay2022transformermemorydifferentiablesearch, mehta2023dsiupdatingtransformermemory}. 
A prominent line of work trains generative retrievers to autoregressively generate item IDs from user context~\cite{tang2023semanticenhanceddifferentiablesearchindex, nguyen2023generativeretrievaldenseretrieval, sun2023learningtokenizegenerativeretrieval, wang2023neuralcorpusindexerdocument}, while subsequent work focuses on scaling and production-oriented training and inference recipes~\cite{pradeep2023doesgenerativeretrievalscale}.

\textbf{Semantic IDs and item tokenization.}
A central component of generative retrieval is \emph{item tokenization}, which maps each item to a short sequence of discrete tokens (Semantic ID) for generation. \textsc{TIGER}~\cite{rajput2023recommendersystemsgenerativeretrieval} popularized learned Semantic IDs via residual quantization (e.g., RQ-VAE~\cite{lee2022autoregressiveimagegenerationusing}) over item content embeddings.
Subsequent works focus on improving SID quality~\cite{wang2025learnableitemtokenizationgenerative, Wang_2024, tang2024generativeretrievalmeetsmultigraded, vandenhirtz2026multimodalgenerativerecommendationfusing, si2024generativeretrievalsemantictreestructured, wu2024higengenerativeretrievallargescale} and show that SID construction is often decisive for downstream retrieval performance.
In particular, \textsc{LETTER}~\cite{wang2025learnableitemtokenizationgenerative} introduces interaction-informed tokenization by adding collaborative regularization to the quantizer training objective. 
However, existing tokenization methods typically assume a static SID vocabulary and do not explicitly address how to preserve compatibility when SIDs are refreshed over time. In contrast, we focus on the \emph{SID update step} under temporal drift: refreshing interaction-informed SIDs while reusing an existing retriever checkpoint.

\textbf{Industry applications of Semantic IDs.}
Several industrial systems report applying Semantic IDs at scale~\cite{yang2024unifyinggenerativedenseretrieval, lin2025unifiedsemanticidrepresentation, yang2025gsidgenerativesemanticindexing}, combining SID tokenization with large generative backbones and production constraints (e.g., constrained decoding and efficient serving).
Representative examples include \textsc{PLUM}~\cite{he2025plumadaptingpretrainedlanguage} and \textsc{OneRec}~\cite{deng2025onerecunifyingretrieverank, zhou2025onerectechnicalreport, zhou2025onerecv2technicalreport}, and Spotify's work on unified generative search and recommendation~\cite{penha2024bridgingsearchrecommendationgenerative}.
While these works demonstrate practical value at scale, they largely focus on SID construction and training/serving recipes, leaving \emph{checkpoint-compatible SID refresh and continual adaptation under temporal drift} underexplored.

\section{Proposed Approach}
\label{sec:method}

We study how to refresh \emph{Semantic IDs (SIDs)} in generative retrieval for sequential recommendation while preserving compatibility with an existing checkpoint.
In real-world deployment scenarios, SID-based retrievers are often finetuned on fresh logs while keeping SIDs fixed, which can make interaction-informed identifiers stale as collaborative patterns drift over time~\cite{he2025plumadaptingpretrainedlanguage, deng2025onerecunifyingretrieverank}.
Our goal is to incorporate fresh interaction signals into SIDs while keeping the deployed retriever warm-startable, avoiding a full rebuild-and-retrain pipeline.

\subsection{Problem Setup}
\label{sec:problem_setup}

We consider the retrieval (candidate generation) stage in sequential recommendation.
Let $\mathcal{U}$ and $\mathcal{I}$ denote the sets of users and items. Each user $u\in\mathcal{U}$ is represented by a time-ordered interaction sequence $S_u = (i_{u,1}, \dots, i_{u,|S_u|})$.
Given $S_u$, retrieval returns a \emph{candidate set} of items (typically hundreds to thousands), deferring fine-grained ordering to a downstream ranker.

We focus on a SID-based generative retriever, where each item $i\in\mathcal{I}$ is assigned a SID represented as a token sequence of length L,
\begin{equation}
\mathbf{z}_i = (z_{i,1}, \dots, z_{i,L}), \quad z_{i,\ell}\in\mathcal{V}_{\ell},
\end{equation}
where $\mathcal{V}_{\ell}$ is the vocabulary of the $\ell$-th codebook.
Conditioned on user context, the retriever generates SID tokens and retrieves candidate items via a SID-to-item mapping.

Crucially, the retriever is tied to the SID space via token embeddings and distributions over $\{\mathcal{V}_\ell\}_{\ell=1}^L$.
Thus, rebuilding SIDs reassigns tokens and makes checkpoint reuse non-trivial. Without special handling, warm-start finetuning can be unstable and is often avoided in favor of retraining from scratch.
Conversely, keeping SIDs fixed while finetuning on fresh interactions can lead to \emph{stale identifiers} that no longer reflect the latest collaborative structure.
Accordingly, we focus on SID update procedures that inject fresh signals from new logs while preserving the original token space, so the retriever can be warm-started rather than fully retrained.

\subsection{Semantic-ID Alignment Update}
\label{sec:method_alignment}

We assume two SID assignments over a largely shared item space:
(i) an \emph{old} SID token assignment produced previously, and
(ii) a \emph{new} SID token assignment rebuilt from fresher logs.
For each item $i$,
\begin{equation}
\mathbf{z}^{\text{old}}_i = (z^{\text{old}}_{i,1},\dots,z^{\text{old}}_{i,L}), \qquad
\mathbf{z}^{\text{new}}_i = (z^{\text{new}}_{i,1},\dots,z^{\text{new}}_{i,L}).
\end{equation}
Our goal is to construct a \emph{bijective function} $\phi$ that aligns the \emph{new} SIDs into the \emph{old} token space, producing \emph{updated} identifiers
\begin{equation}
\tilde{\mathbf{z}}_i = (\tilde z_{i,1},\dots,\tilde z_{i,L}), \qquad
\tilde z_{i,\ell} = \phi_\ell\!\left(z^{\text{new}}_{i,\ell}\right),
\label{eq:update_rule}
\end{equation}
where each $\phi_\ell$ is a one-to-one mapping of tokens \emph{within the same codebook position} (codebook-consistent alignment).
Intuitively, $\tilde{\mathbf{z}}_i$ preserves the token space expected by the existing model, while reflecting updated collaborative structure captured by $\mathbf{z}^{\text{new}}_i$.

To construct $\phi$, we use items that appear in both the old and new tokenizations.
Let $\mathcal{I}_\cap$ denote this overlap set.
For each position $\ell$, we compute co-occurrence weights between new tokens $a$ and old tokens $b$:
\begin{equation}
W_\ell(a,b) \;=\; \sum_{i\in\mathcal{I}_\cap}
\mathbb{I}\!\left[z^{\text{new}}_{i,\ell}=a\right]\cdot
\mathbb{I}\!\left[z^{\text{old}}_{i,\ell}=b\right].
\label{eq:cooc_matrix}
\end{equation}

We then solve a bipartite matching problem between \emph{active} new and old 
tokens at position~$\ell$ (i.e., SID tokens observed in items from $\mathcal{I}_\cap$), where larger values of $W_\ell(a,b)$ indicate more stable correspondences. 
Any one-to-one assignment solver can be used to obtain $\phi_\ell$. In this work, we consider both a lightweight \emph{Greedy} procedure and the \emph{Hungarian} algorithm as possible options, with the choice of solver examined in Section~\ref{sec:experiments}. 
After matching, some new tokens may remain unmapped due to limited overlap 
or sparse usage.
We complete $\phi_\ell$ for all tokens used by the new assignment $\mathbf{z}^{\text{new}}$ by pairing remaining new tokens with remaining (unassigned) old tokens in $\mathcal{V}_\ell$.
Finally, we rewrite all new SID sequences using Eq.~\eqref{eq:update_rule}, obtaining updated SIDs $\tilde{\mathbf{z}}$ that can be directly consumed by the existing retriever checkpoint for warm-start finetuning, without retraining from scratch.

\begin{figure*}[t]
  \centering
  \includegraphics[width=0.9\textwidth]{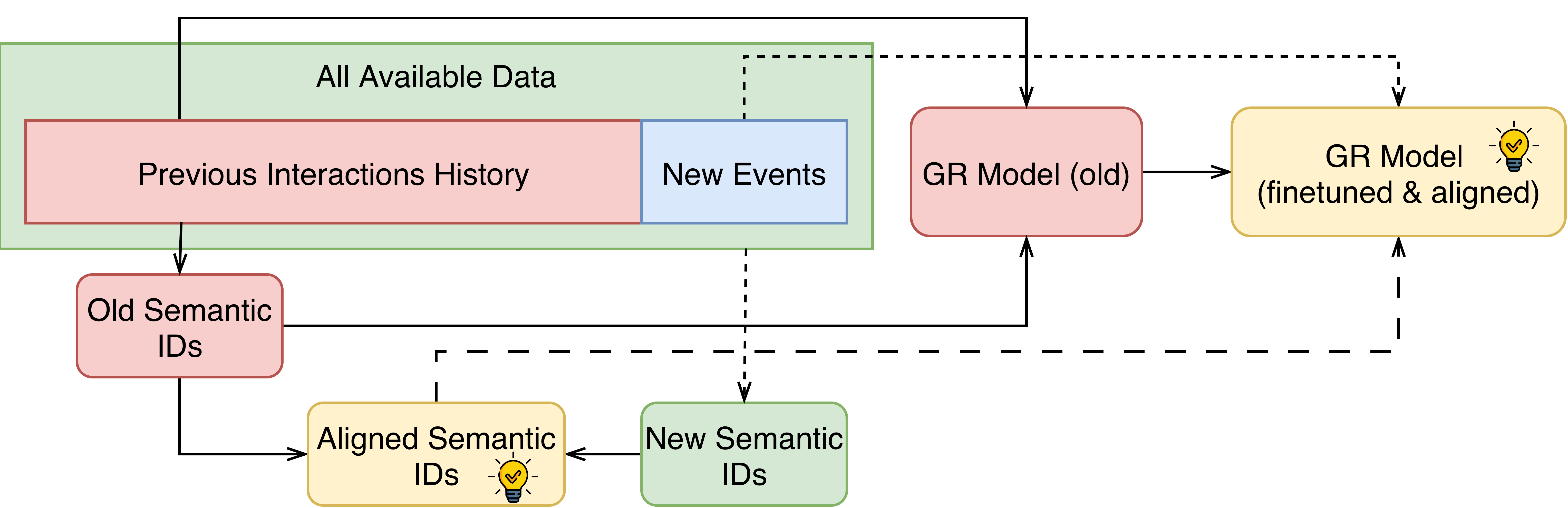}
  \Description{Overview of the proposed training pipeline.}
  \caption{Overview of the proposed training pipeline. The available interaction data are split into two parts: \emph{Previous Interaction History}, used for training the previous model, and \emph{New Events}, used for finetuning. To construct \emph{Aligned SIDs}, we first rebuild SIDs using \emph{All Available Data} and then apply the proposed matching procedure to align them to the previous token space. The downstream model (Generative Retriever) is then finetuned on the \emph{New Events}, warm-starting from the previous checkpoint and using the \emph{Aligned SIDs}. Blocks that are subject to direct improvement are marked with a light bulb sign.}
  \label{fig:overall_pipeline}
\end{figure*}

An overview of the overall alignment \& finetune pipeline is shown in Figure~\ref{fig:overall_pipeline}.

\section{Experiments}
\label{sec:experiments}

In this section, we conduct a systematic evaluation to answer the following research questions:

\begin{itemize}
    \item \textbf{RQ1:} Does alignment enable warm-start finetuning after refreshing interaction-informed SIDs under temporal drift?
    \item \textbf{RQ2:} Over successive refresh-and-finetune steps, does performance remain stable or accumulate degradation?
    \item \textbf{RQ3:} What is the quality--compute trade-off of each policy, in terms of training compute (FLOPs) and alignment overhead?
\end{itemize}

\subsection{Experiment Setup}

\subsubsection{Datasets.}
We evaluate SIDs alignment policies on three public user--item benchmarks spanning e-commerce and streaming.
We use \textsc{Amazon Beauty}~\cite{He_2016} as a standard e-commerce dataset in sequential recommendation.
We further evaluate on \textsc{VK-LSVD}~\cite{poslavsky2026vklsvdlargescaleindustrialdataset}, a short-video streaming benchmark.
To cover a different streaming domain, we include a music streaming dataset \textsc{Yambda}~\cite{ploshkin2025yambda5blargescalemultimodal}.

Following common practice, we apply \emph{5-core} filtering to all datasets.
Table~\ref{tab:dataset_stats} reports resulting statistics.
\begin{table}[t]
\centering
\caption{Dataset statistics after preprocessing.}
\label{tab:dataset_stats}
\scriptsize
\resizebox{\columnwidth}{!}{%
\begin{tabular}{l|rrrr}
\toprule
\textbf{Dataset} & \textbf{\#Users} & \textbf{\#Items} & \textbf{\#Interactions} & \textbf{Density (\%)} \\
\midrule
\textbf{Amazon Beauty} & 22,363 & 12,101 & 198,502 & 0.07 \\
\textbf{VK-LSVD}\footnotemark[1]       & 49,691 & 21,928 & 1,284,340 & 0.12 \\
\textbf{Yambda}\footnotemark[2]        & 6,633 & 33,029 & 631,227 & 0.29 \\
\bottomrule
\end{tabular}
}
\end{table}

\footnotetext[1]{We use the released \texttt{ur0.01\_ir0.01} subsample (1\% random users and items).}
\footnotetext[2]{We use the released \texttt{500M-like} subsample.}

\subsubsection{Baselines.}
We compare two SID paradigms: \textsc{TIGER}-style with content-only tokenization and \textsc{LETTER}-style with interaction-informed tokenization. 
To ensure a controlled comparison, all methods share the same generative retriever architecture and differ only in SID assignment.
Both variants rely on the same RQ-VAE quantizer algorithm.
Different from the original \textsc{LETTER} work, our implementation uses purely collaborative representations based on SASRec embeddings.
We adopt a decoder-only retriever for both paradigms, rather than the encoder--decoder architecture used in the original \textsc{TIGER}. 
We make this choice deliberately to leverage autoregressive training, which reduces training FLOPs per epoch.

For each SID paradigm, we first train a \textbf{Base} model on an earlier time window with a fixed SID vocabulary. This checkpoint is used for all warm-started adaptations.
We then compare three SID update policies: keeping stale SIDs (\textbf{FT-old}), rebuilding SIDs from fresh logs (\textbf{FT-new}), and our checkpoint-compatible alignment update (\textbf{FT-ours}).
We also report \textbf{Full}, which rebuilds SIDs and trains the retriever from scratch, serving as a compute-intensive reference.

\subsubsection{Evaluation protocol.}
We evaluate candidate generation using Recall@K and nDCG@K metrics. Since retrieval stage is mainly responsible for providing a sufficiently large candidate pool for a downstream ranker stage, we prioritize high-recall cutoffs and report $K\in\{10,100,500\}$ (capped at 500 for computational efficiency). \emph{Recall@500 serves as the main metric}.

To avoid temporal leakage from leave-one-out evaluation~\cite{dataleakage, meng2020exploringdatasplittingstrategies, Gusak_2025}, we follow a strict chronological setup: we sort interactions by timestamp and split them into 10 contiguous blocks with an equal number of interactions. We train a \emph{Base} model on blocks~1--8 and use block~9 for model selection (early stopping / epoch budget). Using selected epoch budget, we train the \emph{Full} baseline on blocks~1--9 from scratch. All \emph{FT-*} policies are warm-started from the \emph{Base} checkpoint and use a fixed finetuning budget selected \emph{independently} via a pilot update (train on blocks~1--7, finetune on block~8, select on block~9) and kept fixed across policies and datasets for a fair quality--compute comparison. This configuration intentionally targets short-horizon drift, mirroring typical industrial pipelines where finetuning is triggered frequently and the temporal gap between successive updates is relatively small.

Evaluation is performed on block~10 as a future-item-prediction task. For each user, we construct the inference context using only the recent interaction history available before block~10, i.e., the last $N$ interactions observed in blocks~1--9. The model then retrieves a top-$K$ candidate set, and items interacted with by the user during block~10 are treated as positive targets. This protocol is intended to approximate the deployment setting of industrial retrieval systems (offline scenario), where retriever operates only on historical user behavior available at inference time, while future interactions are naturally unobserved. At the same time, since evaluation is carried out on a chronologically held-out future block, it provides a measure of temporal generalization: the model must transfer from past interaction patterns to long-term user behavior under distribution shift. For a controlled comparison, we restrict histories and evaluation targets to items that are represented in the \emph{old} SID vocabulary, ensuring all policies are evaluated on the same item subset (including \emph{FT-old}, which cannot represent newly-created items without old SIDs). This is a conservative choice: allowing additional items introduced only in the refreshed tokenization could further improve absolute quality. We also experimented with relaxing this constraint and found that relative method performance and ranking remained consistent, suggesting that the observed patterns are not an artifact of SID vocabulary restrictions.

\subsubsection{Implementation Details.}
\label{sec:impl_details}
We construct SIDs using an RQ-VAE quantizer over item embeddings. For \textsc{VK-LSVD} and \textsc{Yambda}, content embeddings are provided with the datasets. For \textsc{Amazon Beauty}, we generate item embeddings using the same encoder and preprocessing pipeline as in~\cite{wang2025learnableitemtokenizationgenerative}. For interaction-informed tokenization, we similarly follow~\cite{wang2025learnableitemtokenizationgenerative} and use trained SASRec collaborative item embeddings. \textsc{TIGER} and \textsc{LETTER} tokenization are implemented in a unified codebase, ensuring consistent retriever and quantizer architectures across variants. The baselines thus reflect key modeling choices of the original systems while abstracting away system-specific engineering.

For a generative retriever architecture we use a decoder-only Transformer with embedding dimension 128, 4 layers, 6 attention heads, feed-forward dimension 1024, ReLU activation, and dropout 0.1. For SID construction, we use an RQ-VAE tokenizer with 4 codebooks of size 512, yielding SID length 4. Both the tokenizer and the downstream retriever are trained with AdamW with learning rate $10^{-4}$ and batch size 256.

Unless stated otherwise, these hyperparameters are shared across datasets. The main dataset-specific choices are the input sequence length and the dimensionality of the item representations used for tokenization. We set the maximum input length to 20 for \textsc{Amazon Beauty} and to 100 for \textsc{Yambda}/\textsc{VK-LSVD}, reflecting the longer interaction histories in the streaming datasets. The RQ-VAE input dimensionality is matched to the dimensionality of the corresponding item embeddings for each dataset. Inference uses constrained beam search restricted to valid existing SIDs with beam width 500. We do not filter out previously interacted items from the model output, as such filtering is typically applied after the retrieval stage in production. For adaptation, we warm-start from the previous checkpoint, keep the learning rate fixed, and reinitialize the optimizer. We report mean metrics over 10 seeds. Significance is assessed on Recall@500 via a paired two-sided Wilcoxon test ($p<0.05$). Additional reproducibility details are provided in the code implementation.\footnotemark[3] 

\footnotetext[3]{Code: \url{https://github.com/iskbaga/semantic-id-alignment}}

\subsection{Overall Performance (RQ1)}
\label{sec:rq1}

\begin{table*}[t]
\centering
\caption{Rolling-origin results on block~10. \textsc{FT-old}/\textsc{FT-new}/\textsc{FT-ours} differ only in SID update policy during warm-start adaptation. \textsc{FT-ours} uses \textsc{Greedy}/\textsc{Hungarian} alignment. Best/2nd-best are bold/underlined.}
\label{tab:overall_performance}
\scriptsize
\renewcommand{\arraystretch}{1.2}   
\resizebox{\textwidth}{!}{%
\begin{tabular}{l|ccc ccc|ccc ccc|ccc ccc}
\toprule
\multirow{3}{*}{\textbf{Model}}
& \multicolumn{6}{c|}{\textbf{Beauty}}
& \multicolumn{6}{c|}{\textbf{Yambda}}
& \multicolumn{6}{c}{\textbf{VK-LSVD}} \\

& \multicolumn{3}{c}{\textbf{Recall}} & \multicolumn{3}{c|}{\textbf{nDCG}}
& \multicolumn{3}{c}{\textbf{Recall}} & \multicolumn{3}{c|}{\textbf{nDCG}}
& \multicolumn{3}{c}{\textbf{Recall}} & \multicolumn{3}{c}{\textbf{nDCG}} \\

& \multicolumn{3}{c}{@10\quad @100\quad @500}
& \multicolumn{3}{c|}{@10\quad @100\quad @500}
& \multicolumn{3}{c}{@10\quad @100\quad @500}
& \multicolumn{3}{c|}{@10\quad @100\quad @500}
& \multicolumn{3}{c}{@10\quad @100\quad @500}
& \multicolumn{3}{c}{@10\quad @100\quad @500} \\

\midrule

\textbf{TIGER} (Base)
& 0.0049 & 0.0276 & 0.0885 & 0.0033 & 0.0087 & 0.0186
& 0.0112 & 0.0394 & 0.0735 & 0.0090 & 0.0175 & 0.0259
& 0.0156 & 0.0866 & 0.2171 & 0.0103 & 0.0291 & 0.0532 \\

\textbf{TIGER} (FT)
& 0.0109 & 0.0659 & 0.1401 & 0.0070 & 0.0205 & 0.0326
& 0.0128 & 0.0479 & 0.0883 & 0.0111 & 0.0222 & 0.0326
& 0.0268 & 0.1404 & 0.2956 & 0.0182 & 0.0491 & 0.0784 \\


\textbf{LETTER} (Base)
& 0.0074 & 0.0440 & 0.1206 & 0.0045 & 0.0133 & 0.0259
& 0.0110 & 0.0378 & 0.0964 & 0.0093 & 0.0173 & 0.0312
& 0.0180 & 0.0994 & 0.2424 & 0.0119 & 0.0334 & 0.0599 \\


\textbf{LETTER} (FT-old)
& 0.0150 & 0.0757 & 0.1645 & 0.0107 & 0.0255 & 0.0402
& 0.0141 & 0.0450 & 0.1116 & 0.0112 & 0.0204 & 0.0362
& \textbf{0.0432} & \textbf{0.1980} & 0.3512 & 0.0281 & \underline{0.0702} & 0.0993 \\

\textbf{LETTER} (FT-new)
& 0.0151 & 0.0876 & 0.1573 & 0.0108 & 0.0285 & 0.0402
& 0.0131 & 0.0505 & 0.1100 & 0.0112 & 0.0226 & 0.0374
& 0.0358 & 0.1780 & 0.3951 & 0.0242 & 0.0622 & 0.1020 \\

\midrule

\textbf{LETTER} (FT-ours, Greedy)
& 0.0152 & 0.0880 & 0.1736 & 0.0104 & 0.0300 & 0.0434
& 0.0134 & \underline{0.0514} & \underline{0.1136} & 0.0114 & \underline{0.0235} & 0.0388
& 0.0412 & \underline{0.1923} & \textbf{0.4098} & \underline{0.0282} & 0.0687 & \textbf{0.1085} \\

\textbf{LETTER} (FT-ours, Hungarian)
& \underline{0.0175} & \underline{0.0892} & \underline{0.1756} & \underline{0.0125} & \underline{0.0310} & \underline{0.0449}
& \textbf{0.0144} & \textbf{0.0519} & \textbf{0.1174} & \underline{0.0125} & \textbf{0.0242} & \textbf{0.0402}
& 0.0394 & 0.1864 & \underline{0.4044} & 0.0273 & 0.0666 & 0.1065  \\

\midrule

\textbf{LETTER} (Full)
& \textbf{0.0254} & \textbf{0.0910} & \textbf{0.1992} & \textbf{0.0180} & \textbf{0.0342} & \textbf{0.0517}
& \underline{0.0142} & 0.0456 & 0.1131 & \textbf{0.0132} & 0.0224 & \underline{0.0392}
& \underline{0.0422} & 0.1908 & 0.3910 & \textbf{0.0296} & \textbf{0.0703} & \underline{0.1080} \\

\bottomrule
\end{tabular}
}
\end{table*}

Table~\ref{tab:overall_performance} compares content-only and interaction-informed SIDs under rolling-origin evaluation (i.e., \textsc{TIGER}-style vs.\ \textsc{LETTER}-style tokenization).
\textsc{Base} models are trained on blocks~1--8.
Warm-start policies (\textsc{FT-*}) finetune the \textsc{Base} checkpoint on block~9 and differ only in how SIDs are handled during this update (\textsc{old/new/ours}).
\textsc{Full} is an expensive reference that rebuilds SIDs and trains the retriever from scratch on blocks~1--9.
All methods follow the same retriever architecture and training recipe.
For \textsc{FT-ours}, we also ablate the alignment solver, using either \textsc{Greedy} matching or the \textsc{Hungarian} algorithm.

\begin{itemize}
    \item \textbf{Interaction-informed SIDs help.}
    \textsc{LETTER} variants generally outperform \textsc{TIGER} across datasets, showing the benefit of injecting collaborative structure into SIDs.

    \item \textbf{Alignment makes SID refresh consistently beneficial under warm-start.}
    Without alignment, rebuilding SIDs (\textsc{FT-new}) is not reliably better than keeping stale SIDs (\textsc{FT-old}). For instance, Recall@500 decreases on Beauty (0.1573 vs.\ 0.1645) and Yambda (0.1100 vs.\ 0.1116).
    This suggests that an unaligned refresh can hinder warm-start adaptation: token reassignment changes the decoding space, so the model must re-learn the mapping during finetuning and may underperform the stale-SID setting.
    In contrast, \textsc{FT-ours} consistently improves high-cutoff retrieval over \textsc{FT-old} on all datasets, e.g., Recall@500 is 0.1756 vs.\ 0.1645 (Beauty), 0.1174 vs.\ 0.1116 (Yambda), and 0.4098 vs.\ 0.3512 (VK-LSVD).

    \item \textbf{\textsc{FT-ours} can be competitive with full retraining.}
    At high cutoffs, \textsc{FT-ours} matches or exceeds \textsc{Full},  (Recall@500: 0.4098 vs.\ 0.3910 on \textsc{VK-LSVD}), while \textsc{Full} remains best on Beauty, suggesting that the optimal refresh strategy may depend on the drift dynamics of the domain.

    \item \textbf{Greedy vs.\ Hungarian.} 
    \textsc{Hungarian} is generally stronger (best on \textsc{Beauty} and \textsc{Yambda}), but greedy wins on \textsc{VK-LSVD}, making it a reasonable low-cost alternative. We assume \textsc{Greedy} approaches or even exceeds \textsc{Hungarian} as number of new block interactions increases.
\end{itemize}

\subsection{Temporal Adaptation (RQ2)}
\label{sec:rq2}

In this section, we study whether SID adaptation is not only effective but also sustainable under repeated updates.
Beyond one-step finetuning (RQ1), we run multi-step continual adaptation on the \textsc{VK-LSVD} short-video streaming benchmark.

We initialize at $t{=}5$ by constructing \textsc{LETTER}-style SIDs and training on blocks~1--5.
For $t\in\{6,7,8\}$, we adapt on block~$t$ and evaluate on block~$t{+}1$.

\emph{Full} constructs SIDs and trains from scratch on blocks~1--$t$.
\emph{FT-old} finetunes on block~$t$ with SIDs fixed from blocks~1--5.
\emph{FT-new} also finetunes on each new block~$t$ but \emph{does} construct fresh SIDs including this new block.
\emph{FT-ours} constructs SIDs at each step, iteratively aligns them to the previous token space, and finetunes on block~$t$. In this experiment, we use \textsc{Greedy} alignment.

\begin{figure}[t]
  \centering
  \includegraphics[width=1.0\linewidth]{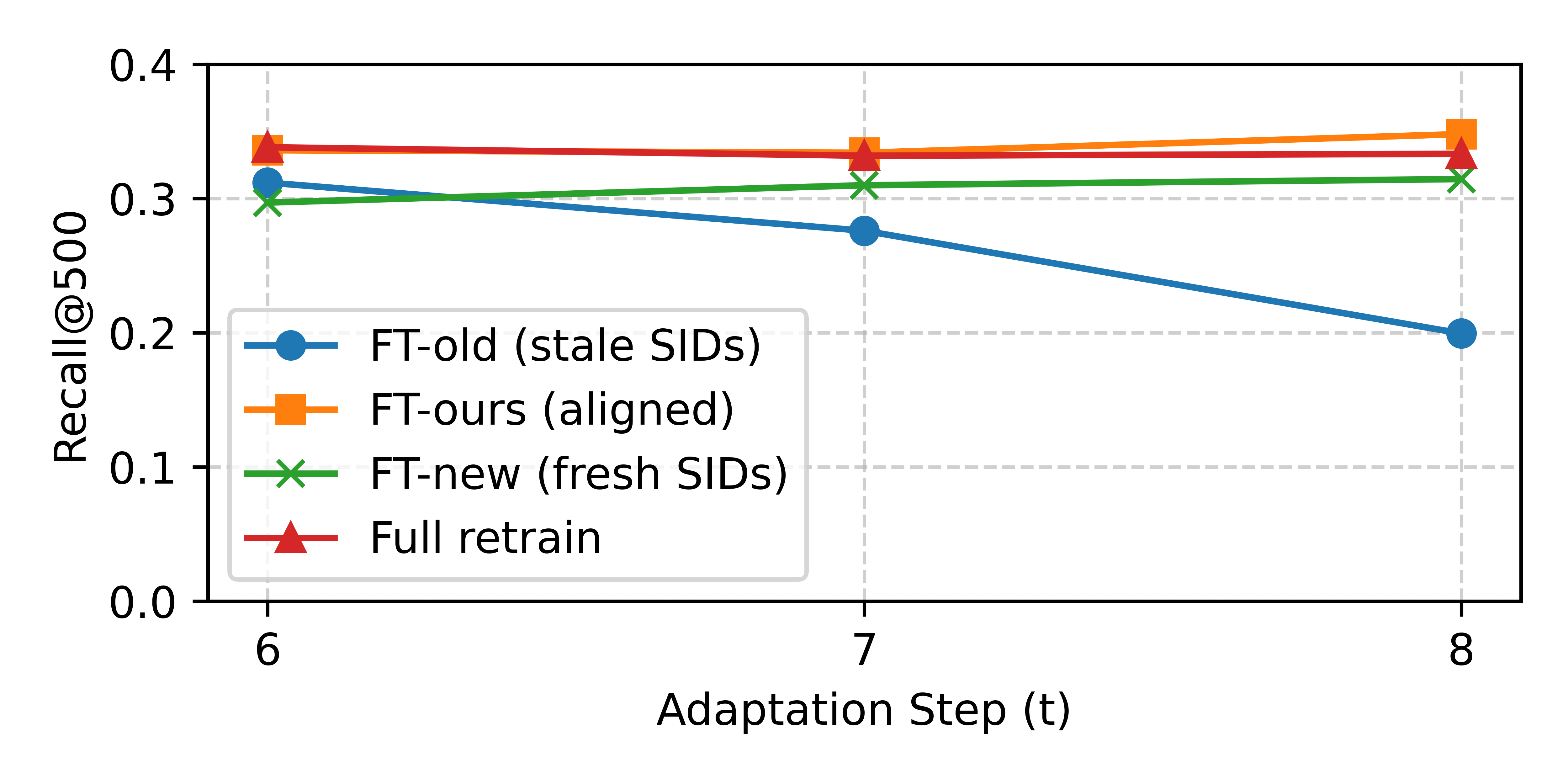}
  \Description{Line plot showing Recall@500 over adaptation steps on VK-LSVD. FT-old declines across steps, FT-ours remains stable and surpasses Full retraining at the final step. FT-new sligtly drops compared to Full }
  \caption{Recall@500 across adaptation steps $t$ on \textsc{VK-LSVD}.}
  \label{fig:temporal_adaptation}
\end{figure}

Figure~\ref{fig:temporal_adaptation} illustrates the Recall@500 results across adaptation steps. \emph{FT-old} suffers a steep performance decline with successive updates. Simply rebuilding the vocabulary at each step (\emph{FT-new}) mitigates this degradation: by the final update it achieves substantially higher recall than \emph{FT-old}. \emph{FT-ours} not only consistently outperforms \emph{FT-old} at every step, but also maintains a high recall level on par with \emph{Full}. These results support our claim that a checkpoint-compatible SID refresh via alignment enables robust continual adaptation under temporal drift, without the need for full retraining.

\subsection{Quality--Compute Trade-off (RQ3)}
\label{sec:rq3}

Beyond retrieval quality, adaptation policies differ substantially in compute cost under repeated updates.
We therefore report a quality--compute trade-off on \textsc{VK-LSVD} using test Recall@500 and training compute in FLOPs, estimated with \texttt{torch.profiler}.

We report two measures: (i) \textbf{$\Delta$TFLOPs}, the additional compute to adapt from the \textsc{Base} checkpoint, and (ii) \textbf{Total TFLOPs}, the cumulative compute for a update scenario (for \textsc{Full}, this includes obtaining \textsc{Base} plus retraining from scratch on blocks~1--9).
For methods that rebuild SIDs (\textsc{FT-new}, \textsc{FT-ours}, \textsc{Full}), FLOPs include retraining the tokenizer (quantizer) in addition to retriever training. Alignment is performed on CPU. Its overhead is analyzed below.

Let $V$ be the codebook size, $L$ the SID length, and $M$ the number of observed co-occurrence pairs per position ($M < V^2$).
Greedy alignment sorts pairs in $O(M\log M)$, while Hungarian runs in $O(V^3)$ per position, $O(LV^3)$ overall.
In practice, codebooks are small ($V=512$, $L=4$), so both are lightweight relative to GPU retriever training.

\begin{table}[t]
\centering
\caption{Quality--compute trade-off for \textsc{LETTER} on \textsc{VK-LSVD}.}
\label{tab:flops_tradeoff}
\scriptsize
\resizebox{\columnwidth}{!}{%
\begin{tabular}{l|c c c}
\toprule
\textbf{Policy} & \textbf{$\Delta$TFLOPs} $\downarrow$ & \textbf{Total TFLOPs} $\downarrow$ & \textbf{R@500} $\uparrow$ \\
\midrule
\textbf{Base}    & --     &  22,975 & 0.2424 \\
\textbf{FT-old}  & 2,815  & 25,790 & 0.3512 \\
\textbf{FT-new}  & 3,003  & 25,978 & 0.3951 \\
\textbf{Full}    & 25,283 & 48,258 & 0.3910 \\
\midrule
\textbf{FT-ours, Greedy} & 3,003$^{\dagger}$ & 25,978$^{\dagger}$ & 0.4098 \\
\bottomrule
\end{tabular}
}

{\footnotesize $^{\dagger}$Our alignment step is CPU-only and introduces negligible FLOPs compared to full retriever pipeline training. We exclude it from FLOPs and report its overhead and time complexity separately.}

\end{table}

As shown in Table~\ref{tab:flops_tradeoff}, warm-start adaptation is far cheaper than full retraining:
\textsc{FT-old} and \textsc{FT-new}/\textsc{FT-ours} require only 2.8--3.0k TFLOPs, yielding 8.97$\times$ and 8.41$\times$ lower compute than \textsc{Full}, respectively.
On \textsc{VK-LSVD}, rebuilding SIDs without alignment already improves Recall@500 over stale SIDs (\textsc{FT-new}: 0.3951 vs.\ \textsc{FT-old}: 0.3512) for a small extra cost (+188 TFLOPs).
Finally, alignment improves the quality--compute frontier: \textsc{FT-ours} matches \textsc{FT-new} in training compute but achieves higher Recall@500 (0.4098 vs.\ 0.3951), while preserving compatibility with the previous SID vocabulary.

As shown in Table~\ref{tab:flops_tradeoff}, the dominant source of computational savings is warm-start finetuning rather than the alignment step itself. In particular, \textsc{FT-new} and \textsc{FT-ours} incur nearly identical training FLOPs, because both require rebuilding SIDs and subsequently finetuning the retriever from the previous checkpoint. The benefit of alignment is instead in improving checkpoint compatibility after SID refresh: it reduces the semantic drift between consecutive SID vocabularies and helps the model adapt to refreshed tokenization while better preserving previously learned structure. As a result, \textsc{FT-ours} achieves higher Recall@500 than \textsc{FT-new} at essentially the same compute cost, thus improving the quality--compute trade-off. We therefore view alignment not as a direct compute-saving component, but as a lightweight mechanism that makes refreshed semantics easier to exploit within a warm-start adaptation pipeline. This property may be particularly useful in setups with more frequent model updates, where controlling semantic drift across successive SID refreshes could make repeated finetuning more stable and potentially reduce the optimization effort required after each update.

\section{Conclusion} 
In this work we propose a lightweight SID alignment method to address identifier staleness in generative retrieval under temporal drift. By preserving token-space compatibility, our approach enables efficient warm-start finetuning on fresh interaction logs. Across three benchmarks, it reduces training compute by 8--9 times while improving high-cutoff recall. Future work includes an exploration of dynamic vocabulary expansion.

\begin{acks}
This research is financially supported by the Foundation for National Technology Initiative's Projects Support as a part of the roadmap implementation for the development of the high-tech field of Artificial Intelligence for the period up to 2030 (agreement 70-2021-00187)
\end{acks}

\bibliographystyle{ACM-Reference-Format}
\balance
\bibliography{references}

\end{document}